\documentclass[11pt]{article}

\usepackage{graphicx,amsmath,amsfonts,amssymb,bm,hyperref,url,breakurl,epsfig,epsf,color,fullpage,MnSymbol,mathbbol,fmtcount,algorithmic,algorithm,semtrans,cite,caption,subcaption,multirow}
  
  \usepackage{mathtools}

\usepackage{titlesec}

\usepackage{tikz}
\usepackage{pgfplots}
\usetikzlibrary{pgfplots.groupplots}

\titleformat{\paragraph}
{\normalfont\normalsize\bfseries}{\theparagraph}{1em}{}
\titlespacing*{\paragraph}
{0pt}{3.25ex plus 1ex minus .2ex}{1.5ex plus .2ex}

\usepackage{movie15}

\usepackage{cite}
\usepackage{caption}
\usepackage[bottom,hang,flushmargin]{footmisc} 

\usepackage{hyperref}
\definecolor{darkred}{RGB}{150,0,0}
\definecolor{darkgreen}{RGB}{0,150,0}
\definecolor{darkblue}{RGB}{0,0,200}
\hypersetup{colorlinks=true, linkcolor=darkred, citecolor=darkgreen, urlcolor=darkblue}

\newtheorem{theorem}{Theorem}[section]
\newtheorem{lemma}[theorem]{Lemma}

\newtheorem{definition}[theorem]{Definition}

\newcommand{\eps}{\varepsilon}

\newcommand{\beq}{\begin{equation}}
\newcommand{\eeq}{\end{equation}}

\newcommand{\nn}{\nonumber}

\newcommand{\A}{{\mtx{A}}}

\newcommand{\z}{{\mtx{z}}}
\newcommand{\dist}{\text{dist}}

\newcommand{\vb}{\mtx{v}}

\newcommand{\li}{\left<}
\newcommand{\ri}{\right>}

\newcommand{\Tc}{\mathcal{T}}
\newcommand{\tn}[1]{\left\|#1\right\|_{\ell_2}}

\newcommand{\zeronorm}[1]{\left\|#1 \right\|_{\ell_0}}

\newcommand{\opnorm}[1]{\left\|#1\right\|}

\newcommand{\twonorm}[1]{\left\|#1\right\|_{\ell_2}}

\newcommand{\abs}[1]{\left|#1\right|}

\newcommand{\x}{\vct{x}}

\definecolor{emmanuel}{RGB}{255,127,0}

\newcommand{\R}{\mathbb{R}}

\newcommand{\E}{\operatorname{\mathbb{E}}}

\newcommand{\vct}[1]{\bm{#1}}
\newcommand{\mtx}[1]{\bm{#1}}

\numberwithin{equation}{section} 

\def \endprf{\hfill {\vrule height6pt width6pt depth0pt}\medskip}

\newcommand*\samethanks[1][\value{footnote}]{\footnotemark[#1]}

\title{Isometric sketching of any set via the \\
Restricted Isometry Property} 

\author{Samet Oymak\thanks{Department of Electrical Engineering and Computer Science, UC Berkeley, Berkeley CA}~\thanks{Simons Institute for the Theory of Computing, UC Berkeley, Berkeley CA}  \quad  Benjamin Recht\samethanks[1]~\thanks{Department of Statistics, UC Berkeley, Berkeley CA} \quad Mahdi
  Soltanolkotabi\thanks{Ming Hsieh Department of Electrical Engineering, University of Southern California, Los Angeles, CA} }
  
\date{June 11, 2015; Revised October 2015}

\begin{document}
\maketitle

\begin{abstract}

In this paper we show that for the purposes of dimensionality reduction certain class of structured random matrices behave similarly to random Gaussian matrices. This class includes several matrices for which matrix-vector multiply can be computed in log-linear time, providing efficient dimensionality reduction of general sets. In particular, we show that using such matrices any set from high dimensions can be embedded into lower dimensions with near optimal distortion. We obtain our results by connecting dimensionality reduction of any set to dimensionality reduction of sparse vectors via a chaining argument. 

\end{abstract}

\section{Introduction}
\emph{Dimensionality reduction} or \emph{sketching} is the problem of embedding a set from high-dimensions into a low-dimensional space, while preserving certain properties of the original high-dimensional set. Such low-dimensional embeddings have found numerous applications in a wide variety of applied and theoretical disciplines across science and engineering. 

Perhaps the most fundamental and popular result for dimensionality reduction is the Johnson-Lindenstrauss (JL) lemma. This lemma states that any set of of $p$ points in high dimensions can be embedded into $\mathcal{O}(\frac{\log p}{\delta^2})$ dimensions, while preserving the Euclidean norm of all points within a multiplicative factor between $1-\delta$ and $1+\delta$. The Johnson-Lindenstrauss Lemma in its modern form can be stated as follows.

\begin{lemma}[Johnson-Lindenstrauss Lemma \cite{johnson1984extensions}]\label{JLlem} Let $\delta\in(0,1)$ and let $\vct{x}_1,\vct{x}_2,\ldots,\vct{x}_p\in\R^n$ be arbitrary points. Then as long as $m=\mathcal{O}(\frac{\log p}{\delta^2})$ there exists a matrix $\mtx{A}\in\R^{m\times n}$ such that
\begin{align}
\label{JLeq}
(1-\delta)\twonorm{\vct{x}_i}\le\twonorm{\mtx{A}\vct{x}_i}\le(1+\delta)\twonorm{\vct{x}_i},
\end{align}
for all $i=1,2,\ldots,p$.
\end{lemma}
This lemma was originally proven to hold with high probability for a matrix $\mtx{A}$ that projects all data points onto a random subspace of dimension $m$ and then scales them by $\sqrt{\frac{n}{m}}$. The result was later generalized so that $\mtx{A}$ could have i.i.d.~normal random entries as well as other random ensembles \cite{dasgupta2003elementary, frankl1988johnson}. More recently the focus has been on constructions of the matrix $\mtx{A}$ where multiplication by this matrix can be implemented efficiently in terms of time and storage e.g.~matrices where it takes at most $o(n\log n)$ time to implement the multiplication. Please see the constructions in \cite{liberty2011dense, ailon2013almost, krahmer2011new, kane2010derandomized, do2009fast} as well as the more recent papers \cite{nelson2014new,ailon2014fast} for further details on related and improved constructions. 

In many uses of dimensionality reduction such as those arising in statistical learning, optimization, numerical linear algebra, etc.~embedding a finite set of points is often not sufficient and one aims to embed a set containing an infinite continuum of points into lower dimensions while preserving the Euclidean norm of all point up to a multiplicative distortion. A classical result due to Gordon \cite{Gor} characterizes the precise tradeoff between distortion, ``size" of the set and the amount of reduction in dimension for a subset of the unit sphere. Before stating this result we need the definition of the Gaussian width of a set which provides a measure of the ``complexity" or ``size" of a set $\mathcal{T}$.
\begin{definition}For a set $\mathcal{T}\subset\R^n$, the mean width $\omega(\mathcal{T})$ is defined as
\begin{align*}
\omega(\mathcal{T})=\E[\sup_{\vct{v}\in\mathcal{T}}\vct{g}^T\vct{v}].
\end{align*}
Here, $\vct{g}\in\R^n$ a Gaussian random vector distributed as $\mathcal{N}(\vct{0},\mtx{I}_n)$.
\end{definition}
\begin{theorem}[Gordon's escape through the mesh] Let $\delta\in(0,1)$, $\mathcal{T}\subset\R^n$ be a subset of the unit sphere ($\mathcal{T}\subset\mathbb{S}^{n-1}$) and let $\mtx{A}\in\R^{m\times n}$ be a matrix with i.i.d $\mathcal{N}(0,1/m)$ entries.\footnote{We note that the factor $1/m$ in the above result is approximate. For the precise result one should replace $1/m$ with $\frac{1}{2}{\left(\frac{\Gamma\left(\frac{m}{2}\right)}{\Gamma\left(\frac{m+1}{2}\right)}\right)}^2\approx 1/m$ where $\Gamma$ denotes the Gamma function.} Then, 
\begin{align}
\label{Gisometry}
\abs{\twonorm{\mtx{A}\vct{x}}-\twonorm{\vct{x}}}\le \delta\twonorm{\vct{x}},
\end{align}
holds for all $\x\in\Tc$ with probability at least $1-2e^{-\frac{\eta^2}{2}}$ as long as
\begin{align}
\label{nummeas}
m\ge\frac{\left(\omega(\mathcal{T})+\eta\right)^2}{\delta^2}.
\end{align}
\end{theorem}
We note that the Johnson-Lindenstrauss lemma for Gaussian matrices follows as a special case. Indeed, for a set $\mathcal{T}$ containing a finite number of points $\abs{\mathcal{T}}\le p$, one can show that $\omega(\mathcal{T})\le\sqrt{2\log p}$ so that the minimal amount of dimension reduction $m$ allowed by \eqref{nummeas} is of the same order as Lemma \ref{JLlem}.

More recently a line of research by Mendelson and collaborators \cite{klartag2005empirical, mendelson2007reconstruction, Mendel1,Mendel2} show that the inequality \eqref{Gisometry} continues to hold for matrices with i.i.d.~sub-Gaussian entries (albeit at a loss in terms of the constants). Please also see \cite{dirksen2014dimensionality,TroppConvex} for more recent results and applications. Connected to this, Bourgain, Dirksen, and Nelson \cite{bourgaintoward} have shown that a similar result to Gordon's theorem continues to hold for certain ensembles of matrices with sparse entries. However, compared to Gordon's result above the allowed reduction in dimension is smaller by constant and logarithmic factors and an additional factor that characterizes the ``spikiness" of the set $\mathcal{T}$.


This paper develops an analogue of Gordon's result for more structured matrices particularly those that have computationally efficient multiplication. At the heart of our analysis is a theorem that shows that matrices that preserve the Euclidean norm of sparse vectors (a.k.a.~RIP matrices), when multiplied by a random sign pattern preserve the Euclidean norm of any set. Roughly stated, linear transforms that provide low distortion embedding of sparse vectors also allow low distortion embedding of any set! We believe that our result provides a rigorous justification for replacing ``slow" Gaussian matrices with ``fast" and computationally friendly matrices in many scientific and engineering disciplines. Indeed, in a companion paper \cite{companion} we utilize our results in this paper to develop sharp rates of convergence for various optimization problems involving such matrices. 

\section{Isometric sketching of sparse vectors}
To connect isometric sketching of sparse vectors to isometric sketching of general sets, we begin by defining the Restricted Isometry Property (RIP). Roughly stated, RIP ensures that a matrix preserves the Euclidean norm of sparse vectors up to a multiplicative distortion $\delta$. This definition immediately implies that RIP matrices can be utilized for isometric sketching of sparse vectors.
\begin{definition}[Restricted Isometry Property] A matrix $\mtx{A}\in\R^{m\times n}$ satisfies the Restricted Isometry Property with distortion $\delta>0$ at a sparsity level $s$, if for all vectors $\vct{x}$ with sparsity at most $s$, we have
\begin{align}
\label{RIPeq}
\abs{ \tn{\mtx{A}\x}^2-\twonorm{\vct{x}}^2}\leq \max(\delta,\delta^2)\tn{\x}^2.
\end{align}
We shall use the short-hand RIP$(\delta,s)$ to denote this property.
\end{definition}
This definition is essentially identical to the classical definition of RIP \cite{candes-tao}. The only difference is that we did not restrict $\delta$ to lie in the interval $[0,1]$. As a result, the correct dependence on $\delta$ in the right-hand side of \eqref{RIPeq} is in the form of $\max(\delta,\delta^2)$. For the purposes of this paper we need a more refined notion of RIP. More specifically, we need RIP to simultaneously hold for different sparsity and distortion levels.
\begin{definition}[Multiresolution RIP] Let $L=\lceil\log_2 n\rceil$. Given $\delta>0$ and a number $s\geq 1$, for $\ell=0,1,2,\ldots,L$, let $(\delta_\ell,s_\ell)=(2^{\ell/2} \delta,2^\ell s)$ be a sequence of distortion and sparsity levels. We say a matrix $\A\in\R^{m\times n}$ satisfies the Multiresolution Restricted Isometry Property (MRIP) with distortion $\delta>0$ at sparsity $s$, if for all $\ell\in\{1,2,\ldots,L\}$, RIP($\delta_\ell,s_\ell$) holds. More precisely for vectors of sparsity at most $s_\ell$ ($\zeronorm{\vct{x}}\le s_\ell$) the sequence of inequalities 
\begin{align}
\label{MRIPeq}
\abs{ \tn{\mtx{A}\x}^2-\twonorm{\vct{x}}^2}\leq \max(\delta_\ell,\delta_\ell^2)\tn{\x}^2,
\end{align}
simultaneously holds for all $\ell\in\{1,2,\ldots,L\}$. We shall use the short-hand MRIP$(\delta,s)$ to denote this property.
\end{definition}
This definition essentially requires the matrix to satisfy RIP at different scales. At the lowest scale, it reduces to the standard RIP$(\delta,s)$ definition. Noting that $s_L=2^Ls\ge n$ at the highest scale this condition requires
\begin{align*}
\abs{ \tn{\mtx{A}\x}^2-\twonorm{\vct{x}}^2}\leq \max(\delta_L,\delta_L^2)\tn{\x}^2,
\end{align*}
to hold for all vectors $\vct{x}\in\R^n$. While this condition looks rather abstract at first sight, with proper scaling it can be easily satisfied for popular random matrix ensembles used for dimensionality reduction.

\section{From isometric sketching of sparse vectors to general sets}
Our main result states that a matrix obeying Multiresolution RIP with the right distortion level $\tilde{\delta}$ can be used for embedding any subset $\Tc$ of $\R^n$.
\begin{theorem} \label{rip prop} Let $\mathcal{T}\subset\R^n$ and suppose the matrix $\mtx{H}\in\R^{m\times n}$ obeys the Multiresolution RIP with sparsity and distortion levels 
\begin{align}
\label{levels}
s=150(1+\eta)\quad\text{and}\quad \tilde{\delta}= \frac{\delta\cdot\emph{rad}(\mathcal{T})}{C\max\left(\emph{rad}(\mathcal{T}),\omega(\mathcal{T})\right)},
\end{align}
with $C>0$ an absolute constant. Then, for a diagonal matrix $\mtx{D}$ with an i.i.d.~random sign pattern on the diagonal, the matrix $\mtx{A}=\mtx{H}\mtx{D}$ obeys
\begin{align}
\label{gordontype}
\sup_{\vct{x}\in\mathcal{T}}|\tn{\A\x}^2-\tn{\x}^2|\leq \max\left(\delta,\delta^2\right)\cdot\left(\emph{rad}(\mathcal{T})\right)^2,
\end{align}
with probability at least $1-\exp(-\eta)$. Here, $\emph{rad}(\mathcal{T})=\sup_{\vct{v}\in\mathcal{T}} \twonorm{\vct{v}}$ is the maximum Euclidean norm of a point inside $\mathcal{T}$.
\end{theorem}
This theorem shows that given a matrix that is good for isometric embedding of sparse vectors when multiplying its columns by a random sign pattern it becomes suitable for isometric embedding of any set! For typical random matrix ensembles that are commonly used for dimensionality reduction purposes, given a sparsity $s$ and distortion $\tilde{\delta}$ the minimum dimension $m$ for the MRIP$(s,\tilde{\delta})$ to hold grows as $m\sim \frac{s}{\tilde{\delta}^2}$. In Theorem \ref{rip prop}, we have $s\sim1$ and $\tilde{\delta}\sim\frac{\delta}{\omega(\mathcal{T})}$ so that the minimum dimension $m$ for \eqref{gordontype} to hold is of the order of $m\sim \frac{\omega^2(\mathcal{T})}{\delta^2}$. This is exactly the same scaling one would obtain by using Gaussian random matrices via Gordon's lemma in \eqref{nummeas}. To see this more clearly we now focus on applying Theorem \ref{rip prop} to random matrices obtained by subsampling a unitary matrix.
\begin{definition}[Subsampled Orthogonal with Random Sign (SORS) matrices]\label{SOSdef} Let $\mtx{F}\in\R^{n\times n}$ denote an orthonormal matrix obeying
\begin{align}
\label{BOS}
\mtx{F}^*\mtx{F}=\mtx{I}\quad\text{and}\quad\max_{i,j}\abs{\mtx{F}_{ij}}\le \frac{\Delta}{\sqrt{n}}.
\end{align}
Define the random subsampled matrix $\mtx{H}\in\R^{m\times n}$ with i.i.d. rows chosen uniformly at random from the rows of $\mtx{F}$. Now we define the Subsampled Orthogonal with Random Sign (SORS) measurement ensemble as $\mtx{A}=\mtx{H}\mtx{D}$, where $\mtx{D}\in\R^{n\times n}$ is a random diagonal matrix with the diagonal entries i.i.d.~$\pm 1$ with equal probability.
\end{definition}
To simplify exposition, in the definition above we have focused on SORS matrices based on subsampled orthonormal matrices $\mtx{H}$ with i.i.d.~rows chosen uniformly at random from the rows of an orthonormal matrix $\mtx{F}$ obeying \eqref{BOS}. However, our results continue to hold for SORS matrices defined via a much broader class of random matrices $\mtx{H}$ with i.i.d.~rows chosen according to a probability measure on Bounded Orthonormal Systems (BOS). Please see \cite[Section 12.1]{foucart2013random} for further details on such ensembles.
By utilizing results on Restricted Isometry Property of subsampled orthogonal random matrices obeying \eqref{BOS} we can show that the Multi-resolution RIP holds at the sparsity and distortion levels required by \eqref{levels}. Therefore, Theorem \ref{rip prop} immediately implies a result similar to Gordon's lemma for SORS matrices.

\begin{theorem} \label{SOSthm} Let $\mathcal{T}\subset\R^n$ and suppose $\A\in\R^{m\times n}$ is selected from the SORS distribution of Definition \ref{SOSdef}. Then, 
\begin{align}
\label{gordontypeSOS}
\sup_{\vct{x}\in\mathcal{T}}|\tn{\A\x}^2-\tn{\x}^2|\leq \max\{\delta,\delta^2\}\cdot \left(\emph{rad}(\mathcal{T})\right)^2,
\end{align}
holds with probability at least $1-2e^{-\eta}$ as long as
\begin{align}
\label{samplesSOS}
m\geq C\Delta^2(1+\eta)^2(\log n)^4\text{ }\frac{\max\left(1,\frac{\omega^2(\mathcal{T})}{\left(\emph{rad}(\mathcal{T})\right)^2}\right)}{\delta^2}.\end{align}
\end{theorem}
As we mentioned earlier while we have stated the result for real valued SORS matrices obeying \eqref{BOS}, the result can be generalized to complex matrices and more broadly to SORS matrices obtained from Bounded Orthonormal Systems. We would also like to point out that one can improve the dependence on $\eta$ and potentially replace a few $\log n$ factors with $\log\left(\omega(\mathcal{T})\right)$ by utilizing improved RIP bounds such as \cite{RudelSparse, cheraghchi2013restricted, dirksen2013tail}. We note that any future result that reduces log factors in the sample complexity of RIP will also automatically improve the lower bound on $m$ in our results. Infact, after the first version of this manuscript became available there has been a very interesting reduction of log factors by Haviv and Regev in \cite{haviv2015restricted}. We believe that utilizing this new RIP result it may be possible to improve the lower bound in \eqref{samplesSOS} to 
\begin{align}
\label{improvlb}
m\geq C\Delta^2(1+\eta)^2(\log \omega(\mathcal{T}))^2\log n\text{ }\frac{\max\left(1,\frac{\omega^2(\mathcal{T})}{\left(\emph{rad}(\mathcal{T})\right)^2}\right)}{\delta^2}.
\end{align}
We leave this for future research.\footnote{The reason \eqref{improvlb} does not follow immediately from the results in  \cite{haviv2015restricted} is twofold: (1) The results of \cite{haviv2015restricted} are based on more classical definitions of RIP (without the $\max(\delta,\delta^2)$ as in \eqref{RIPeq}) and (2) the dependence on the distortion level $\delta$ in terms of sample complexity is not of the form $1/\delta^2$ and has slightly weaker dependence of the form $\frac{\log^4(1/\delta)}{\delta^2}$ which holds for sufficiently small $\delta$.}
%

Ignoring constant/logarithmic factors Theorem \ref{SOSthm} is an exact analogue of Gordon's lemma for Gaussian matrices in terms of the tradeoff between the reduced dimension $m$ and the distortion level $\delta$. Gordon's result for Gaussian matrices has been utilized in numerous problems. Theorem \ref{SOSthm} above allows one to replace Gaussian matrices with SORS matrices for such problems. For example, Chandrasekaran et al. \cite{Cha} use Gordon's lemma to obtain near optimal sample complexity bounds for linear inverse problems involving Gaussian matrices. An immediate application of Theorem \ref{SOSthm} implies near optimal sample complexity results using SORS matrices. To the extent of our knowledge this is the first sample optimal result using a computational friendly matrix. We refer the reader to our companion paper for further detail \cite{companion}.


Theorem \ref{SOSthm} is the first result to establish an analogue to Gordon's Theorem that holds for all sets $\mathcal{T}$, while using matrices that have fast multiplication. We would like to pause however to mention a few interesting results that hold with additional assumptions on the set $\mathcal{T}$. Perhaps, the first results of this kind were established for the Restricted Isometry Property in \cite{candes-tao, RudelSparse}, where the set $\mathcal{T}$ is the set of vectors with a certain sparsity level. In \cite{krahmer2011new} Krahmer and Ward established a JL type embedding for RIP matrices with columns multiplied by a random sign pattern. That is, the authors show that Theorem \ref{SOSthm} holds when $\mathcal{T}$ is a finite point cloud. More recently, in \cite{yap2013stable} the authors show a Gordon type embedding result holds for manifold signals using RIP matrices whose columns are multiplied by a random sign pattern. Earlier, we mentioned the very interesting result of Bourgain et.~al.~\cite{bourgaintoward} which establishes a result in the spirit of Theorem \ref{SOSthm} for sparse matrices. However, compared with \eqref{samplesSOS}, the minimum dimension $m$ \cite{bourgaintoward}, in addition to the mean width $\omega(\mathcal{T})$ and distortion $\delta$, also depends on a parameter which characterizes the spikiness of the set $\mathcal{T}$. This is of course to be expected as when using sparse ensembles it is not possible to embed spiky sets into lower dimension without significant loss in terms of distortion. In addition, the authors of \cite{bourgaintoward} also establish results without the spikiness assumption for particular $\mathcal{T}$ using Fast Johnson-Lindenstrauss (FJLT) matrices e.g.~see \cite[Section 6.2]{bourgaintoward}. Recently, Pilanci and Wainwright in \cite{pilanci2014randomized} have established a result of similar flavor to Theorem \ref{SOSthm} but with suboptimal tradeoff between the allowed dimension reduction and the complexity of the set $\mathcal{T}$. Roughly stated, this result requires $m\gtrsim \left(\log n\right)^4\frac{\omega^4(\mathcal{T})}{\delta^2}$ using a sub-sampled Hadamard matrix combined with a diagonal matrix of i.i.d.~Rademacher random variables.\footnote{We would like to point out that our proofs also hint at an alternative proof strategy to that of \cite{pilanci2014randomized} if one is interested in establishing $m\gtrsim(\log n)^4\frac{\omega^4(\mathcal{T})}{\delta^2}$. In particular, one can cover the set $\mathcal{T}$ with Euclidean balls of size $\delta$. Based on Sudakov's inequality the logarithm of the size of this cover is at most $\frac{\omega^2(\mathcal{T})}{\delta^2}$. One can then relate this cover to a cover obtained by using a random pseudo-metric such as the one defined in \cite{RudelSparse}. As a result one incurs an additional factor $(\log n)^4\omega^2(\mathcal{T})$. Multiplying these two factors leads to the requirement $m\gtrsim (\log n)^4\frac{\omega^4(\mathcal{T})}{\delta^2}$. }

%

\section{Proofs}
Before we move to the proof of the main theorem we begin by stating known results on RIP for bounded orthogonal systems and show how Theorem \ref{SOSthm} follows from our main theorem (Theorem \ref{rip prop}).

\subsection{Proof of Theorem \ref{SOSthm} for SORS matrices}
We first state a classical result on RIP originally due to Rudelson and Vershynin \cite{RudelSparse,Vers}. We state the version in \cite{foucart2013random} which holds generally for bounded orthogonal systems. We remark that the results in \cite{RudelSparse,Vers} as well as those of \cite{foucart2013random} are stated for the regime $\delta<1$. However, by going through the analysis of these papers carefully one can confirm that our definition of RIP (with max$(\delta,\delta^2)$ on the right-hand side in lieu of $\delta$) continues to hold for $\delta\geq 1$. 
\begin{lemma} [RIP for sparse signals, \cite{RudelSparse,Vers,foucart2013random}] \label{rip for sparse}Let $\mtx{F}\in\R^{n\times n}$ denote an orthonormal matrix obeying
\begin{align}
\label{BOS2}
\mtx{F}^*\mtx{F}=\mtx{I}\quad\text{and}\quad\max_{i,j}\abs{\mtx{F}_{ij}}\le \frac{\Delta}{\sqrt{n}}.
\end{align}
Define the random subsampled matrix $\mtx{H}\in\R^{m\times n}$ with i.i.d. rows chosen uniformly at random from the rows of $\mtx{F}$. Then RIP$(\delta,s)$ holds with probability at least $1-e^{-\eta}$ for all $\delta>0$ as long as
\begin{align*}
m\ge C\Delta^2\frac{s\left(\log^3 n\log m+\eta\right)}{\delta^2}.
\end{align*}
Here $C>0$ is a fixed numerical constant.
\end{lemma}
Applying the union bound over $L=\lceil \log n\rceil$ sparsity levels and using the change of variable $\eta\rightarrow \eta+\log L$, together with the fact that $(\log n)^4+\eta\le (1+\eta)(\log n)^4$, Lemma \ref{rip for sparse} immediately leads to the following lemma.
\begin{lemma} Consider $\mtx{H}\in\R^{m\times n}$ distributed as in Lemma \ref{rip for sparse}. $\mtx{H}$ obeys multi-resolution RIP with sparsity $s$ and distortion $\tilde{\delta}>0$ with probability $1-e^{-\eta}$ as long as
\beq
m\geq C(1+\eta)\Delta^2\frac{s(\log n)^4}{\tilde{\delta}^2}\nn.
\eeq
\end{lemma}
Theorem \ref{SOSthm} now follows by using $s=C(1+\eta)$ and $\tilde{\delta}=\frac{\delta}{C\max\left(1,\frac{\omega(\mathcal{T})}{\text{rad}(\mathcal{T})}\right)}$ in Theorem \ref{rip prop}.

\subsection{Connection between JL-embedding and RIP}
\label{JL2RIP}
A critical tool in our proof is an interesting result due to Krahmer and Ward \cite{krahmer2011new} which shows that RIP matrices with columns multiplied by a random sign pattern obey the JL lemma.
\begin{theorem} [Discrete JL embedding via RIP, \cite{krahmer2011new}] \label{wk thm} Assume $\mathcal{T}\subset\R^n$ is a finite set of points. Suppose $\mtx{H}\in\R^{m\times n}$ is a matrix satisfying RIP$(s,\delta)$ with sparsity $s$ and distortion $\delta>0$ obeying 
\begin{align*}
s\le\min\left(40(\log\left(4|\mathcal{T}|\right)+\eta),n\right)\quad\text{and}\quad 0<\delta\leq \frac{\eps}{4},
\end{align*}
where $\mtx{D}\in\R^{n\times n}$ is a random diagonal matrix with the diagonal entries i.i.d.~$\pm 1$ with equal probability. Then the matrix $\mtx{A}=\mtx{H}\mtx{D}$ obeys
\begin{align}
\label{wardeq}
\abs{\tn{\mtx{A}\x}^2-\tn{\x}^2}\leq \max(\eps,\eps^2)\tn{\x}^2,
\end{align}
simultaneously for all $\vct{x}\in\mathcal{T}$ with probability at least $1-e^{-\eta}$.
\end{theorem}
The above theorem differs from the result of Krahmer and Ward \cite{krahmer2011new} in two ways. First, the authors state their result for $0<\eps<1$. Furthermore, in the right-hand side of \eqref{wardeq} the authors use $\epsilon$ in lieu of $\max(\epsilon,\epsilon^2)$. However, it is easy to verify that their proof (with essentially no modifications) can accommodate the result stated above.

\subsection{Generic chaining related notations and definitions}
Our proof makes use of the generic chaining machinery e.g.~see \cite{talagrand2006generic}. We gather some of the required definitions and notations in this section. Define $N_0=1$ and $N_\ell=2^{2^\ell}$ for $\ell\geq 1$.

\begin{definition}[Admissible sequence, \cite{talagrand2006generic}] Given a set $\mathcal{T}$ an admissible sequence is an increasing sequence ($\mathcal{A}_\ell$) of partitions of $\mathcal{T}$ such that $\abs{\mathcal{A}_\ell}\le N_\ell$.
\end{definition}
As noted by Talagrand, increasing sequence of partitions means that every set of $\mathcal{A}_{\ell+1}$ is
contained in a set of $\mathcal{A}_\ell$ and $\mathcal{A}_\ell(t)$ is the unique
element of $\mathcal{A}_\ell$ that contains $t$. Then the $\gamma_2$ functional is defined as
\beq
\gamma_2(\mathcal{T})=\inf\underset{t}{\sup}\sum_{\ell=0}^\infty 2^{\ell/2}\text{rad}(\mathcal{A}_\ell(t)),\nn
\eeq
where the infimum is taken over all admissible sequences. Let $\bar{\mathcal{A}}_\ell$ be one such optimal admissible sequence. Based on this sequence we have we define the successive covers.
\begin{definition}[successive covers]
\label{Tell}
Using $\tilde{A}_\ell$ we construct successive covers $\mathcal{T}_\ell$ of $\mathcal{T}$ by taking the center point point\footnote{The center point of a set is the center of the smallest ball containing that set.} of each set of $\bar{\mathcal{A}}_\ell$. 
\end{definition}
%
%

Let $e_{\ell}(\vb)$ be the associated distortion of the cover with respect to a point $\vb$ i.e.~$e_{\ell}(\vb)=\dist(\vb,\mathcal{T}_\ell)$. Then for all $\vb\in \mathcal{T}$, the $\gamma_2$ functional obeys
\beq
\sum_{\ell=0}^\infty 2^{\ell/2}e_{\ell}(\vb) \leq \gamma_2(\mathcal{T}).\nn
\eeq

It is well known that $\gamma_2(\mathcal{T})$ and Gaussian width $\omega(\mathcal{T})$ are of the same order. More precisely, for a fixed numerical constant $C$
\beq
C^{-1}\omega(\mathcal{T})\leq \gamma_2(\mathcal{T})\leq C\omega(\mathcal{T}).\nn
\eeq
Given the distortion $\delta$ in the statement of Theorem \ref{rip prop} we also define different scales of distortion 
\beq
\delta_0=\delta,~\delta_1=2^{1/2}\delta,~\dots,~\delta_L=2^{L/2}\delta\nn,
\eeq
with $L=\log_2\lceil n\rceil$. 

\subsection{Proof of Theorem \ref{rip prop}}
Without loss of generality we assume that rad$(\mathcal{T})=1$. We begin by noting that the Multi-resolution RIP property combined with the powerful JL-embedding result stated in Theorem \ref{wk thm} allows for JL embedding at different distortion levels. We apply such an argument to successively more refined covers of the set $\mathcal{T}$ and at different distortion scales inside a generic chaining type argument to arrive at the proof for an arbitrary (and potentially continuous) set $\mathcal{T}$. We should point out that one can also follow an alternative approach which leads to the same conclusion. Instead of using multi-resolution RIP, we could have defined a ``multi-resolution embedding property" for the mapping $\A$ that isometrically maps finite set of points $\mathcal{T}$ with a near optimal set cardinality-distortion tradeoff at varying levels. One can show that this property also implies isometric embedding of a continuous set $\mathcal{T}$.
We begin by stating a lemma which shows isometric embedding as well as a few other properties for points belonging to the refined covers $\mathcal{T}_\ell$ at different distortion levels $\delta_\ell$. The proof of this lemma is deferred to Section \ref{laterproof}.
\begin{lemma} \label{lemma obvious}Suppose $\mtx{H}\in\R^{m\times n}$ obeys MRIP$(s,\frac{\delta}{4})$ with distortion level $\delta$ and sparsity $s=150(1+\eta)$. Furthermore, let $\mtx{D}\in\R^{n\times n}$ be a diagonal matrix with a random i.i.d. sign pattern on the diagonal and set $\mtx{A}=\mtx{H}\mtx{D}$. Also let $\mathcal{T}_\ell$ be successive refinements of the set $\mathcal{T}$ from Definition \ref{Tell}.
Then, with probability at least $1-\exp(-\eta)$ the followings identities hold simultaneously for all $\ell=1,2,\ldots, L$,
\begin{itemize}
\item For all $ \vb\in \mathcal{T}_{\ell-1}\cup\mathcal{T}_\ell\cup(\mathcal{T}_{\ell-1}-\mathcal{T}_{\ell})$,
\begin{align}
\label{ith step0}
\tn{\A\vct{v}}\leq (1+2^{\ell/2}\delta)\tn{\vct{v}}.
\end{align}
\item For all $ \vb\in \mathcal{T}_{\ell-1}\cup\mathcal{T}_\ell\cup(\mathcal{T}_{\ell-1}-\mathcal{T}_{\ell})$,
\beq
|\tn{\A\vct{v}}^2-\tn{\vct{v}}^2|\leq \max\left(2^{\ell/2}\delta,2^\ell\delta^2\right)\cdot\twonorm{\vct{v}}^2.\label{ith step}\\
\eeq
\item For all $\vct{u}\in \mathcal{T}_{\ell-1}$ and $\vct{v}\in \mathcal{T}_{\ell}-\{\vct{u}\}$,
\begin{align}
\label{lastone}
\abs{\vct{u}^*\A^*\A\vct{v}-\vct{u}^*\vct{v}}\leq \max\left(2^{\ell/2}\delta,2^\ell\delta^2\right)\cdot\twonorm{\vct{u}}\twonorm{\vct{v}}.
\end{align}
\end{itemize}
\end{lemma}
With this lemma in place we are ready to prove our main theorem. To this aim given a point $\vct{x}\in\mathcal{T}$, for $\ell=1,2,\ldots,L$ let $\vct{z}_\ell$ be the closest neighbor of $\vct{x}$ in $\mathcal{T}_\ell$. We also define $\vct{z}_{L+1}=\vct{x}$. We note that $\vct{z}_\ell$ depends on $\vct{x}$. For ease of presentation we do not make this dependence explicit. We also drop $\x$ from the distortion term $e_\ell(\x)$ and simply use $e_\ell$. Now observe that for all $\ell=1,2,\ldots,L$,  we have
\begin{align}
\label{bnddiffz}
\tn{\z_\ell-\z_{\ell-1}}\leq \tn{\z_\ell-\x}+\tn{\z_{\ell-1}-\x}\leq e_\ell+e_{\ell-1}\leq 2e_{\ell-1}.
\end{align}
We are interested in bounding $|\tn{\A\x}^2-\tn{\x}^2|$ for all $\vct{x}\in\mathcal{T}$. Define $\tilde{L}=\max\left(0,\lfloor2\log_2\left(\frac{1}{\delta}\right)\rfloor\right)$, and note that applying the triangular inequality we have
\begin{align}
\label{mymainx}
|\tn{\A\x}^2-\tn{\x}^2|\le&\abs{\twonorm{\mtx{A}\vct{z}_{\tilde{L}}}^2-\twonorm{\vct{z}_{\tilde{L}}}^2}+\abs{\tn{\A\x}^2-\tn{\A\vct{z}_{\tilde{L}}}^2}+\abs{\twonorm{\vct{x}}^2-\twonorm{\vct{z}_{\tilde{L}}}^2}\nonumber\\
\le&\sum_{\ell=1}^{\tilde{L}}\left(\abs{\twonorm{\mtx{A}\vct{z}_\ell}^2-\twonorm{\vct{z}_\ell}^2}-\abs{\twonorm{\mtx{A}\vct{z}_{\ell-1}}^2-\twonorm{\vct{z}_{\ell-1}}^2}\right)\nonumber\\
&+\abs{\tn{\A\x}^2-\tn{\A\vct{z}_{\tilde{L}}}^2}+\abs{\twonorm{\vct{x}}^2-\twonorm{\vct{z}_{\tilde{L}}}^2}+\abs{\twonorm{\mtx{A}\vct{z}_{0}}^2-\twonorm{\vct{z}_{0}}^2}.
\end{align}
First note that by Lemma \ref{lemma obvious}
\begin{align*}
|\tn{\A\z_0}^2-\tn{\z_0}^2|\leq \max\left(\delta,\delta^2\right)\tn{\z_0}^2\leq \max\left(\delta,\delta^2\right).
\end{align*}
Using the above inequality in \eqref{mymainx} we arrive at 
\begin{align}
\label{mainineq}
|\tn{\A\x}^2-\tn{\x}^2|\le&\sum_{\ell=1}^{\tilde{L}}\left(\abs{\twonorm{\mtx{A}\vct{z}_\ell}^2-\twonorm{\vct{z}_\ell}^2}-\abs{\twonorm{\mtx{A}\vct{z}_{\ell-1}}^2-\twonorm{\vct{z}_{\ell-1}}^2}\right)\nonumber\\
&+\abs{\tn{\A\x}^2-\tn{\A\vct{z}_{\tilde{L}}}^2}+\abs{\twonorm{\vct{x}}^2-\twonorm{\vct{z}_{\tilde{L}}}^2}+\max\left(\delta,\delta^2\right).
\end{align}
We now proceed by bounding each of the first three terms in \eqref{mainineq}. Before getting into the details of these bounds we would like to point out that \eqref{mainineq}, as well as the results presented in Sections \ref{secfirsterm}, \ref{secsecterm} and \ref{secthirdterm} are derived under the assumption that $\tilde{L}\le L$. Proper modification allows us to bound $|\tn{\A\x}^2-\tn{\x}^2|$ even when $\tilde{L}> L$. We shall explain this argument in complete detail in Section \ref{analog59}. 

\subsubsection{Bounding the first term in \eqref{mainineq}}\label{secfirsterm}
For $1\le\ell\le \tilde{L}$, we have $\delta_\ell=2^{\ell/2}\delta\le 1$ so that $\max(\delta_\ell,\delta_\ell^2)=\delta_\ell$. Thus, applying Lemma \ref{lemma obvious} together with \eqref{bnddiffz} we arrive at
\begin{align}
\label{secbndnd}
\abs{\tn{\A(\z_\ell-\z_{\ell-1})}^2-\tn{\z_\ell-\z_{\ell-1}}^2}\leq& 2^{\ell/2}\delta\twonorm{\vct{z}_\ell-\vct{z}_{\ell-1}}^2\le 2^{\ell/2+2}\delta e_{\ell-1}^2,
\end{align}
and
\begin{align}
\label{thirdbndnd}
\abs{\langle\A(\z_\ell-\z_{\ell-1}),\A\z_{\ell-1}\ri-\li\z_\ell-\z_{\ell-1},\z_{\ell-1}\rangle}\leq 2^{\ell/2+1}\delta e_{\ell-1}.
\end{align}
The triangular inequality yields
\begin{align*}
\abs{\tn{\A\z_\ell}^2-\twonorm{\vct{z}_\ell}^2}=&\abs{\tn{\A(\z_\ell-\z_{\ell-1})+\A\z_{\ell-1}}^2-\twonorm{\vct{z}_\ell}^2}\\
\le&\abs{\tn{\A(\z_\ell-\z_{\ell-1})}^2-\twonorm{\vct{z}_\ell-\vct{z}_{\ell-1}}^2}+\abs{\tn{\A\z_{\ell-1}}^2-\twonorm{\vct{z}_{\ell-1}}^2}\\
&+2\abs{\li\A(\z_\ell-\z_{\ell-1}),\A\z_{\ell-1}\ri-\langle\vct{z}_\ell-\vct{z}_{\ell-1},\vct{z}_{\ell-1}\rangle}.
\end{align*}
Combining the latter with \eqref{secbndnd} and \eqref{thirdbndnd} we arrive at the following recursion
\begin{align}
\label{mainrecur}
\abs{\tn{\A\z_\ell}^2-\twonorm{\vct{z}_\ell}^2}-\abs{\twonorm{\mtx{A}\vct{z}_{\ell-1}}^2-\twonorm{\vct{z}_{\ell-1}}^2}\le\delta\left(2e_{\ell-1}+4e_{\ell-1}^2\right)2^{\ell/2}.
\end{align}
Adding both sides of the above inequality for $1\leq \ell\leq \tilde{L}$, and using $e_\ell^2\leq 2e_\ell\leq 4$, we arrive at
\begin{align}
\sum_{\ell=1}^{\tilde{L}}\left(\abs{\twonorm{\mtx{A}\vct{z}_\ell}^2-\twonorm{\vct{z}_\ell}^2}-\abs{\twonorm{\mtx{A}\vct{z}_{\ell-1}}^2-\twonorm{\vct{z}_{\ell-1}}^2}\right)\le&10\delta\left(\sum_{\ell=1}^{\tilde{L}}2^{\ell/2}e_{\ell-1}\right)\nonumber\\
=&10\sqrt{2}\delta\left(\sum_{\ell=0}^{\tilde{L}-1}2^{\ell/2}e_{\ell}\right)\nonumber\\
=&10\sqrt{2}\delta\gamma_2(\mathcal{T}).\label{term1}
\end{align}

\subsubsection{Bounding the second term in \eqref{mainineq}}\label{secsecterm}
To bound the second term we begin by bounding $\abs{\twonorm{\mtx{A}\vct{x}}-\twonorm{\mtx{A}\vct{z}_{\tilde{L}}}}$. To this aim first note that since MRIP$(s,\frac{\delta}{4})$ holds for $\mtx{H}$ with $s=150(1+\eta)$ then $s_L=150\times2^L(1+\eta)\ge n$. As a result for all $\vct{x}\in\R^n$ we have
\begin{align*}
\abs{\twonorm{\mtx{H}\vct{x}}^2-\twonorm{\vct{x}}^2}\le\max(\frac{1}{4}\delta_L,\frac{1}{16}\delta_L^2)\twonorm{\vct{x}}^2.
\end{align*}
Using the simple inequality $1+\max(\delta,\delta^2)\le(1+\delta)^2$, this immediately implies
\begin{align}
\label{firstsimp}
\opnorm{\mtx{A}}=\opnorm{\mtx{H}}\le\frac{1}{4}2^{\frac{L}{2}}\delta+1.
\end{align}
Furthermore, by the definition of $N_\ell$ we have $\twonorm{\vct{x}-\vct{z}_L}\le e_L$. 
These two inequalities together with repeated use of the triangular inequality we have
\begin{align*}
\abs{\twonorm{\mtx{A}\vct{x}}-\twonorm{\mtx{A}\vct{z}_{\tilde{L}}}}=&\abs{\twonorm{\mtx{A}\vct{x}}-\twonorm{\mtx{A}\vct{z}_L}+\twonorm{\mtx{A}\vct{z}_L}-\twonorm{\mtx{A}\vct{z}_{\tilde{L}}}}\\
\le&\twonorm{\mtx{A}(\vct{x}-\vct{z}_L)}+\twonorm{\mtx{A}(\vct{z}_{L}-\vct{z}_{\tilde{L}})}\\
\le&\opnorm{\mtx{A}}\twonorm{\vct{x}-\vct{z}_L}+\twonorm{\sum_{\ell=\tilde{L}+1}^{L}\mtx{A}(\vct{z}_\ell-\vct{z}_{\ell-1})}\\
\le&\left(\frac{1}{4}2^{\frac{L}{2}}\delta+1\right)e_L+\sum_{\ell=\tilde{L}+1}^{L}\twonorm{\mtx{A}(\vct{z}_\ell-\vct{z}_{\ell-1})}.
\end{align*}
Using Lemma \ref{lemma obvious} equation \eqref{ith step0} in the above inequality and noting that for $\ell>\tilde{L}$, we have $2^{\ell/2}\delta\ge 1$ we conclude that
\begin{align}
\label{firstdiff}
\abs{\twonorm{\mtx{A}\vct{x}}-\twonorm{\mtx{A}\vct{z}_{\tilde{L}}}}\le&\left(\frac{1}{4}2^{\frac{L}{2}}\delta+1\right)e_L+\sum_{\ell=\tilde{L}+1}^{L}(1+2^{\ell/2}\delta)\tn{\vct{z}_\ell-\vct{z}_{\ell-1}}\nonumber\\
\le&\frac{5}{4}2^{L/2}\delta e_L+\sum_{\ell=\tilde{L}+1}^{L}2^{\ell/2+1}\delta\tn{\vct{z}_\ell-\vct{z}_{\ell-1}}\nonumber\\
\le&\frac{5}{4}\delta2^{L/2}e_L+4\sqrt{2}\delta\sum_{\ell=\tilde{L}+1}^{L}2^{(\ell-1)/2}e_{\ell-1}\nonumber\\
\le&4\sqrt{2}\delta\left(\sum_{\ell=\tilde{L}}^L 2^{\ell/2}e_\ell\right)\nonumber\\
\le&4\sqrt{2}\delta\gamma_2(\mathcal{T}).
\end{align} 

Now note that by Lemma \ref{lemma obvious} equation \eqref{ith step0} and using the fact that rad$(\mathcal{T})=1$, we know that $\twonorm{\mtx{A}\vct{z}_{\tilde{L}}}\le 1+2^{\tilde{L}/2}\delta\le 2$. Thus, using this inequality together with \eqref{firstdiff} we arrive at
\begin{align}
\label{term2}
\abs{\twonorm{\mtx{A}\vct{x}}^2-\twonorm{\mtx{A}\vct{z}_{\tilde{L}}^2}}\le&\abs{\twonorm{\mtx{A}\vct{x}}-\twonorm{\mtx{A}\vct{z}_{\tilde{L}}}}\abs{\twonorm{\mtx{A}\vct{x}}+\twonorm{\mtx{A}\vct{z}_{\tilde{L}}}}\nonumber\\
\le&\abs{\twonorm{\mtx{A}\vct{x}}-\twonorm{\mtx{A}\vct{z}_{\tilde{L}}}}^2+\abs{\twonorm{\mtx{A}\vct{x}}-\twonorm{\mtx{A}\vct{z}_{\tilde{L}}}}\twonorm{\mtx{A}\vct{z}_{\tilde{L}}}\nonumber\\
\le&32\delta^2\gamma_2^2(\mathcal{T})+8\sqrt{2}\delta\gamma_2(\mathcal{T}).
\end{align}

\subsubsection{Bounding the third term in \eqref{mainineq}}\label{secthirdterm}
Similar to the second term we begin by bounding $\abs{\twonorm{\vct{x}}-\twonorm{\vct{z}_{\tilde{L}}}}$. Noting that $2^{\ell/2}\delta\ge 1$ for $\ell\ge \tilde{L}$ we have
\begin{align*}
\abs{\twonorm{\vct{x}}-\twonorm{\vct{z}_{\tilde{L}}}}\le\sum_{\ell=\tilde{L}}^L\twonorm{\vct{z}_{\ell+1}-\vct{z}_\ell}\le2\sum_{\ell=\tilde{L}}^Le_\ell\le2\sum_{\ell=\tilde{L}}^L2^{\ell/2}\delta e_\ell\le2\delta\gamma_2(\mathcal{T}).
\end{align*}
Thus using this inequality together with the fact that $\twonorm{\vct{z}_{\tilde{L}}}\le 1$ we arrive at
\begin{align}
\label{term3}
\abs{\twonorm{\vct{x}}^2-\twonorm{\vct{z}_{\tilde{L}}}^2}=&\abs{\twonorm{\vct{x}}-\twonorm{\vct{z}_{\tilde{L}}}}\cdot\left(\twonorm{\vct{x}}+\twonorm{\vct{z}_{\tilde{L}}}\right)\nonumber\\
\le&\abs{\twonorm{\vct{x}}-\twonorm{\vct{z}_{\tilde{L}}}}^2+\abs{\twonorm{\vct{x}}-\twonorm{\vct{z}_{\tilde{L}}}}\nonumber\\
\le&4\delta^2\gamma_2^2(\mathcal{T})+2\delta\gamma_2(\mathcal{T}).
\end{align}

\subsubsection{Establishing an analog of \eqref{mainineq} and the bounds \eqref{term1}, \eqref{term2}, and \eqref{term3} when $\tilde{L}>L$}
\label{analog59}
This section describes how an analog of \eqref{mainineq} as well as the subsequent bounds in Sections \ref{secfirsterm}, \ref{secsecterm} and \ref{secthirdterm} can be derived when $\tilde{L}>L$. Using similar arguments leading to the derivation of \eqref{mainineq} we arrive at
\begin{align}
\label{mainineq2}
|\tn{\A\x}^2-\tn{\x}^2|\le&\sum_{\ell=1}^{L}\left(\abs{\twonorm{\mtx{A}\vct{z}_\ell}^2-\twonorm{\vct{z}_\ell}^2}-\abs{\twonorm{\mtx{A}\vct{z}_{\ell-1}}^2-\twonorm{\vct{z}_{\ell-1}}^2}\right)\nonumber\\
&+\abs{\twonorm{\mtx{A}\vct{x}}^2-\twonorm{\vct{x}}^2}-\abs{\twonorm{\mtx{A}\vct{z}_L}^2-\twonorm{\vct{z}_L}^2}+\max\left(\delta,\delta^2\right).
\end{align}
The main difference with the $\tilde{L}\le L$ case is that we let the summation in the first term go upto $L$ and instead of studying the second line of \eqref{mainineq}, we will directly bound the difference $\abs{\tn{\A\x}^2-\tn{\x}^2}-\abs{\tn{\A\z_L}^2-\tn{\z_L}^2}$ in \eqref{mainineq2}. 

We now turn our attention to bounding the first two terms in \eqref{mainineq2}. For the first term in \eqref{mainineq2} an argument identical to the derivation of \eqref{term1} in Section \ref{secfirsterm} allows us to conclude
\begin{align}
\sum_{\ell=1}^{L}\left(\abs{\twonorm{\mtx{A}\vct{z}_\ell}^2-\twonorm{\vct{z}_\ell}^2}-\abs{\twonorm{\mtx{A}\vct{z}_{\ell-1}}^2-\twonorm{\vct{z}_{\ell-1}}^2}\right)\le10\sqrt{2}\delta\gamma_2(\mathcal{T}).\label{term1prim}
\end{align}
To bound the second term in \eqref{mainineq2} note that we have
\begin{align}
\label{quad last line}
\abs{\tn{\A\x}^2-\tn{\x}^2}-\big|&\tn{\A\z_L}^2-\tn{\z_L}^2\big|\nonumber\\
\le&\abs{\left(\tn{\A\x}^2-\tn{\A\z_L}^2\right)-\left(\tn{\x}^2-\tn{\z_L}^2\right)},\nonumber\\
=&\abs{\left(\tn{\A\left(\x-\vct{z}_L\right)+\A\vct{z}_L}^2-\tn{\A\z_L}^2\right)-\left(\tn{(\x-\vct{z}_L)+\vct{z}_L}^2-\tn{\z_L}^2\right)},\nonumber\\
=&\abs{\left(\twonorm{\mtx{A}(\vct{x}-\vct{z}_L)}^2-\twonorm{\vct{x}-\vct{z}_L}^2\right)+2\left(\langle\mtx{A}(\vct{x}-\vct{z}_L),\A\vct{z}_L\rangle-\langle\vct{x}-\vct{z}_L,\vct{z}_L\rangle\right)},\nonumber\\
\le&\abs{\twonorm{\mtx{A}(\vct{x}-\vct{z}_L)}^2-\twonorm{\vct{x}-\vct{z}_L}^2}+2\abs{\langle\mtx{A}(\vct{x}-\vct{z}_L),\A\vct{z}_L\rangle-\langle\vct{x}-\vct{z}_L,\vct{z}_L\rangle},\nonumber\\
=&\abs{\twonorm{\mtx{A}(\vct{x}-\vct{z}_L)}^2-\twonorm{\vct{x}-\vct{z}_L}^2}+2\twonorm{\vct{x}-\vct{z}_L}\abs{\langle\mtx{A}\frac{\vct{x}-\vct{z}_L}{\twonorm{\vct{x}-\vct{z}_L}},\A\vct{z}_L\rangle-\langle\frac{\vct{x}-\vct{z}_L}{\twonorm{\vct{x}-\vct{z}_L}},\vct{z}_L\rangle},\nonumber\\
\le&\abs{\twonorm{\mtx{A}(\vct{x}-\vct{z}_L)}^2-\twonorm{\vct{x}-\vct{z}_L}^2}\nonumber\\
&+\frac{1}{2}\twonorm{\vct{x}-\vct{z}_L}\abs{\twonorm{\mtx{A}\left(\frac{\vct{x}-\vct{z}_L}{\twonorm{\vct{x}-\vct{z}_L}}+\vct{z}_L\right)}^2-\twonorm{\frac{\vct{x}-\vct{z}_L}{\twonorm{\vct{x}-\vct{z}_L}}+\vct{z}_L}^2}\nonumber\\
&+\frac{1}{2}\twonorm{\vct{x}-\vct{z}_L}\abs{\twonorm{\mtx{A}\left(\frac{\vct{x}-\vct{z}_L}{\twonorm{\vct{x}-\vct{z}_L}}-\vct{z}_L\right)}^2-\twonorm{\frac{\vct{x}-\vct{z}_L}{\twonorm{\vct{x}-\vct{z}_L}}-\vct{z}_L}^2}.
\end{align}
To complete our bound note that since MRIP$(s,\frac{\delta}{4})$ holds for $\mtx{A}$ with $s=150(1+\eta)$ then $s_L=150\times2^L(1+\eta)\ge n$. As a result for all $\vct{w}\in\R^n$ we have
\begin{align*}
\abs{\twonorm{\mtx{A}\vct{w}}^2-\twonorm{\vct{w}}^2}\le\max(\frac{1}{4}\delta_L,\frac{1}{16}\delta_L^2)\twonorm{\vct{w}}^2.
\end{align*}
For $\tilde{L}>L$ we have $\delta_L=2^{\frac{L}{2}}\delta\le 1$ which immediately implies that for all $\vct{w}\in\R^n$ we have
\begin{align}
\label{diffA}
\abs{\twonorm{\mtx{A}\vct{w}}^2-\twonorm{\vct{w}}^2}\le\frac{1}{4}2^{L/2}\delta\twonorm{\vct{w}}^2.
\end{align}
Now using \eqref{diffA} with $\vct{w}=\vct{x}-\vct{z}_L, \frac{\vct{x}-\vct{z}_L}{\twonorm{\vct{x}-\vct{z}_L}}-\vct{z}_L$, and $\frac{\vct{x}-\vct{z}_L}{\twonorm{\vct{x}-\vct{z}_L}}+\vct{z}_L$ in \eqref{quad last line} and noting that $\twonorm{\vct{z}_L}\le$rad$(\mathcal{T})\le1$, we conclude that
\begin{align}
\label{myreallygoodt}
\abs{\tn{\A\x}^2-\tn{\x}^2}-\big|\tn{\A\z_L}^2-\tn{\z_L}^2\big|\le&\frac{1}{4}2^{L/2}\delta\twonorm{\vct{x}-\vct{z}_L}^2+\frac{1}{8}2^{L/2}\delta\twonorm{\vct{x}-\vct{z}_L}\twonorm{\frac{\vct{x}-\vct{z}_L}{\twonorm{\vct{x}-\vct{z}_L}}+\vct{z}_L}^2\nonumber\\
&+\frac{1}{8}2^{L/2}\delta\twonorm{\vct{x}-\vct{z}_L}\twonorm{\frac{\vct{x}-\vct{z}_L}{\twonorm{\vct{x}-\vct{z}_L}}-\vct{z}_L}^2\nonumber\\
\le&\frac{1}{4}2^{L/2}\delta\twonorm{\vct{x}-\vct{z}_L}^2+2^{L/2}\delta\twonorm{\vct{x}-\vct{z}_L}\nonumber\\
\le&2^{L/2}\delta \left(\frac{1}{4}e_L^2+e_L\right)\nonumber\\
\le&\frac{3}{2}2^{L/2}\delta e_L\nonumber\\
\le&\frac{3}{2}\delta \gamma_2(\mathcal{T}).
\end{align}
Plugging \eqref{term1prim} and \eqref{myreallygoodt} into \eqref{mainineq2} we arrive at
\begin{align}
\label{mainineq5}
\abs{\tn{\A\x}^2-\tn{\x}^2}\le 16\delta \gamma_2(\mathcal{T})+\max(\delta,\delta^2).
\end{align}
%
%
%
%
\subsubsection{Finishing the proof of Theorem \ref{rip prop}}
To finish off the proof we plug in the bounds from \eqref{term1}, \eqref{term2}, and \eqref{term3} into \eqref{mainineq} and use the fact that $\gamma_2(\mathcal{T})\le C\omega(\mathcal{T})$ for a fixed numerical constant $C$, to conclude that for $\tilde{L}\le L$ we have
\begin{align}
\label{mymainuppb2}
|\tn{\A\x}^2-\tn{\x}^2|\le&10\sqrt{2}\delta\gamma_2(\mathcal{T})+32\delta^2\gamma_2^2(\mathcal{T})+8\sqrt{2}\delta\gamma_2(\mathcal{T})+4\delta^2\gamma_2^2(\mathcal{T})+2\delta\gamma_2(\mathcal{T})+\max(\delta,\delta^2)\nonumber\\
\le&36\delta^2C^2\omega^2(\mathcal{T})+28C\delta\omega(\mathcal{T})+\max(\delta,\delta^2)\nonumber\\
\le&72\cdot\max\left(C\delta\omega(\mathcal{T}),C^2\delta^2\omega^2(\mathcal{T})\right)+\max(\delta,\delta^2)\nonumber\\
\le&73\cdot\max\left(C\delta\left(\max(1,\omega(\mathcal{T}))\right),C^2\delta^2\left(\max(1,\omega(\mathcal{T}))\right)^2\right).
\end{align}
Combining this with the fact that \eqref{mainineq5} holds for $L>\tilde{L}$ we can conclude that for all $\vct{x}\in\mathcal{T}$
\begin{align}
\label{mymainuppb}
|\tn{\A\x}^2-\tn{\x}^2|\le 73\cdot\max\left(C\delta\left(\max(1,\omega(\mathcal{T}))\right),C^2\delta^2\left(\max(1,\omega(\mathcal{T}))\right)^2\right).
\end{align}

Note that assuming MRIP$(s,\frac{\delta}{4})$ with $s=150(1+\eta)$ we have arrived at \eqref{mymainuppb}. Applying the change of variable
\begin{align*}
\delta\rightarrow\frac{\delta}{292C\max\left(1,\omega(\mathcal{T})\right)},
\end{align*}
we can conclude that under the stated assumptions of the theorem for all $\vct{x}\in\mathcal{T}$
\begin{align*}
|\tn{\A\x}^2-\tn{\x}^2|\le\max(\delta,\delta^2),
\end{align*}
completing the proof. Now all that remains is to prove Lemma \ref{lemma obvious}. This is the subject of the next section.
\subsubsection{Proof of Lemma \ref{lemma obvious}}
\label{laterproof}
For a set $\mathcal{M}$ we define the normalized set $\widetilde{\mathcal{M}}=\big\{\frac{\vct{v}}{\twonorm{\vct{v}}}:\text{ }\vct{v}\in\mathcal{M}\big\}$. We shall also define
\begin{align*}
\mathcal{Q}_\ell=\mathcal{T}_{\ell-1}\cup \mathcal{T}_\ell\cup (\mathcal{T}_{\ell}-\mathcal{T}_{\ell-1})\cup \left(\widetilde{(\mathcal{T}_\ell-\mathcal{T}_{\ell-1})}-\widetilde{\mathcal{T}}_{\ell-1}\right)\cup \left(\widetilde{(\mathcal{T}_\ell-\mathcal{T}_{\ell-1})}+\widetilde{\mathcal{T}}_{\ell-1}\right).
\end{align*}
We will first prove that for $\ell=1,2,\ldots,L$ and every $\vct{v}\in\mathcal{Q}_\ell$
\begin{align}
\label{ithstepnew}
|\tn{\A\vct{v}}^2-\tn{\vct{v}}^2|\leq \max\left(2^{\ell/2}\delta,2^\ell\delta^2\right)\cdot\twonorm{\vct{v}}^2,
\end{align}
holds with probability at least $1-e^{-\eta}$. We then explain how the other identities follow from this result. To this aim, note that that by the assumptions of the lemma MRIP$(s,\frac{\delta}{4})$ holds for the matrix $\mtx{H}$ with $s=150(1+\eta)$. By definition this is equivalent to RIP($s_\ell,\delta_\ell)$ holding for $\ell=1,2,\ldots,L$ with $(s_\ell,\frac{\delta_\ell}{4})=(2^\ell s,\frac{2^{\ell/2}\delta}{4})$. Now observe that the number of entries of $\mathcal{Q}_\ell$ obeys $\abs{\mathcal{Q}_\ell}\le5N_\ell^2$ with $N_\ell=2^{2^\ell}$ which implies
\begin{align}
\label{sell}
s_\ell=&2^\ell s\nonumber\\
=&2^\ell\left(150+150\eta\right)\nonumber\\
\ge&2^\ell\left(40(\log 2)(\log_2 (20)+1)+\frac{1}{2}(\eta+1)\right)\nonumber\\
\ge&2^\ell\left(40(\log 2)\left(\frac{\log_2 (20)}{2^\ell}+1\right)+\frac{\ell}{2^\ell}(\eta+1)\right)\nonumber\\
\ge&40(\log 2)\left(\log_2(20)+2^\ell\right)+\ell(\eta+1)\nonumber\\
\ge&40\log\left(4\abs{\mathcal{Q}_\ell}\right)+\ell(\eta+1)\nonumber\\
\ge&\min\left(40\log\left(4\abs{\mathcal{Q}_\ell}\right)+\ell(\eta+1),n\right).
\end{align}
By the MRIP assumption, RIP$(s_\ell,\frac{\delta_\ell}{4})$ holds for $\mtx{H}$. This together with \eqref{sell} allows us to apply Theorem \ref{wk thm} to conclude that for each $\ell=1,2,\ldots,L$ and every $\vct{x}\in\mathcal{Q}_\ell$
\begin{align*}
\abs{\twonorm{\mtx{A}\vct{x}}^2-\twonorm{\vct{x}}^2}\le\max(\delta_\ell,\delta_\ell^2)\twonorm{\vct{x}}^2,
\end{align*}
holds with probability at least $1-e^{-\ell(\eta+1)}$. Noting that
\begin{align*}
\sum_{\ell=1}^L e^{-\ell(\eta+1)}\le\sum_{\ell=1}^\infty e^{-\ell(\eta+1)}=\frac{e^{-(\eta+1)}}{1-e^{-(\eta+1)}}\le e^{-\eta},
\end{align*}
completes the proof of \eqref{ithstepnew} by the union bound. 

We note that since $\mathcal{T}_{\ell-1}\cup\mathcal{T}_\ell\cup(\mathcal{T}_\ell-\mathcal{T}_{\ell-1})\subset\mathcal{Q}_\ell$, \eqref{ithstepnew} immediately implies \eqref{ith step}. The proof of \eqref{ith step0} follows from the proof of \eqref{ith step} by noting that 
\begin{align*}
(1+\delta_\ell)^2\geq 1+\max\left(\delta_\ell,\delta_\ell^2\right).
\end{align*}
To prove \eqref{lastone}, first note that $\frac{\vct{v}}{\twonorm{\vct{v}}}-\frac{\vct{u}}{\twonorm{\vct{u}}}\in \widetilde{(\mathcal{T}_\ell-\mathcal{T}_{\ell-1})}-\widetilde{\mathcal{T}}_{\ell-1}$ and $\frac{\vct{u}}{\twonorm{\vct{u}}}+\frac{\vct{v}}{\twonorm{\vct{v}}}\in \widetilde{(\mathcal{T}_\ell-\mathcal{T}_{\ell-1})}+\widetilde{\mathcal{T}}_{\ell-1}$. Hence, applying \eqref{ithstepnew}
\begin{align*}
\abs{\tn{\A\left(\frac{\vct{u}}{\twonorm{\vct{u}}}+\frac{\vct{v}}{\twonorm{\vct{v}}}\right)}^2-\tn{\frac{\vct{u}}{\twonorm{\vct{u}}}+\frac{\vct{v}}{\twonorm{\vct{v}}}}^2}\leq& \max(\delta_\ell,\delta_\ell^2)\tn{\frac{\vct{u}}{\twonorm{\vct{u}}}+\frac{\vct{v}}{\twonorm{\vct{v}}}}^2\nn\\
\abs{\tn{\A\left(\frac{\vct{v}}{\twonorm{\vct{v}}}-\frac{\vct{u}}{\twonorm{\vct{u}}}\right)}^2-\tn{\frac{\vct{v}}{\twonorm{\vct{v}}}-\frac{\vct{u}}{\twonorm{\vct{u}}}}^2}\leq& \max(\delta_\ell,\delta_\ell^2)\tn{\frac{\vct{v}}{\twonorm{\vct{v}}}-\frac{\vct{u}}{\twonorm{\vct{u}}}}^2.
\end{align*}
Summing these two identities and applying the triangular inequality we conclude that
\begin{align*}
\frac{1}{\twonorm{\vct{u}}\twonorm{\vct{v}}}\abs{\vct{u}^*\A^*\A\vct{v}-\vct{u}^*\vct{v}}\leq \frac{1}{4}\max(\delta_\ell,\delta_\ell^2)\left(\tn{\frac{\vct{u}}{\twonorm{\vct{u}}}+\frac{\vct{v}}{\twonorm{\vct{v}}}}^2+\tn{\frac{\vct{v}}{\twonorm{\vct{v}}}-\frac{\vct{u}}{\twonorm{\vct{u}}}}^2\right)=\max(\delta_\ell,\delta_\ell^2),
\end{align*}
completing the proof of \eqref{lastone}.

\subsection*{Acknowledgements}
BR is generously supported by ONR awards N00014-11-1-0723 and N00014-13-1-0129, NSF awards CCF-1148243 and CCF-1217058, AFOSR award FA9550-13-1-0138, and a Sloan Research Fellowship.  SO was generously supported by the Simons Institute for the Theory of Computing and NSF award CCF-1217058.  This research is supported in part by NSF CISE Expeditions Award CCF-1139158, LBNL Award 7076018, and DARPA XData Award FA8750-12-2-0331, and gifts from Amazon Web Services, Google, SAP, The Thomas and Stacey Siebel Foundation, Adatao, Adobe, Apple, Inc., Blue Goji, Bosch, C3Energy, Cisco, Cray, Cloudera, EMC2, Ericsson, Facebook, Guavus, HP, Huawei, Informatica, Intel, Microsoft, NetApp, Pivotal, Samsung, Schlumberger, Splunk, Virdata and VMware. We thank Ahmed El Alaoui for a careful reading of the manuscript. We also thank Sjoerd Dirksen and Jelani Nelson for helpful comments and also for pointing us to some useful references on generalizing Gordon's result to matrices with sub-Gaussian entries. We would also like to thank Christopher J. Rozell for bringing the paper \cite{yap2013stable} on stable and efficient embedding of manifold signals to our attention.

\bibliography{Bibfiles}

\begin{thebibliography}{10}

\bibitem{ailon2013almost}
N.~Ailon and E.~Liberty.
\newblock An almost optimal unrestricted fast {J}ohnson-{L}indenstrauss
  transform.
\newblock {\em ACM Transactions on Algorithms (TALG)}, 9(3):21, 2013.

\bibitem{ailon2014fast}
N.~Ailon and H.~Rauhut.
\newblock Fast and {RIP}-optimal transforms.
\newblock {\em Discrete \& Computational Geometry}, 52(4):780--798, 2014.

\bibitem{bourgaintoward}
J.~Bourgain, S.~Dirksen, and J.~Nelson.
\newblock Toward a unified theory of sparse dimensionality reduction in
  {E}uclidean space.
\newblock {\em arXiv preprint arXiv:1311.2542}, 2013.

\bibitem{candes-tao}
E.~J. Candes and T.~Tao.
\newblock Decoding by linear programming.
\newblock {\em IEEE Transactions on Information Theory}, 51(12):4203--4215,
  2005.

\bibitem{Cha}
V.~Chandrasekaran, B.~Recht, P.~A. Parrilo, and A.~S. Willsky.
\newblock The convex geometry of linear inverse problems.
\newblock {\em Foundations of Computational Mathematics}, 12(6):805--849, 2012.

\bibitem{cheraghchi2013restricted}
M.~Cheraghchi, V.~Guruswami, and A.~Velingker.
\newblock Restricted isometry of {F}ourier matrices and list decodability of
  random linear codes.
\newblock {\em SIAM Journal on Computing}, 42(5):1888--1914, 2013.

\bibitem{dasgupta2003elementary}
S.~Dasgupta and A.~Gupta.
\newblock An elementary proof of a theorem of {J}ohnson and {L}indenstrauss.
\newblock {\em Random Structures \& Algorithms}, 22(1):60--65, 2003.

\bibitem{dirksen2013tail}
S.~Dirksen.
\newblock Tail bounds via generic chaining.
\newblock {\em arXiv preprint arXiv:1309.3522}, 2013.

\bibitem{dirksen2014dimensionality}
S.~Dirksen.
\newblock Dimensionality reduction with subgaussian matrices: a unified theory.
\newblock {\em arXiv preprint arXiv:1402.3973}, 2014.

\bibitem{do2009fast}
T.~T. Do, L.~Gan, Y.~Chen, N.~Nguyen, and T.~D. Tran.
\newblock Fast and efficient dimensionality reduction using structurally random
  matrices.
\newblock In {\em IEEE International Conference on Acoustics, Speech and Signal
  Processing, ICASSP 2009.}, pages 1821--1824.

\bibitem{foucart2013random}
S.~Foucart and H.~Rauhut.
\newblock Random sampling in bounded orthonormal systems.
\newblock In {\em A Mathematical Introduction to {C}ompressive {S}ensing},
  pages 367--433. Springer, 2013.

\bibitem{frankl1988johnson}
P.~Frankl and H.~Maehara.
\newblock The {J}ohnson-{L}indenstrauss lemma and the sphericity of some
  graphs.
\newblock {\em Journal of Combinatorial Theory, Series B}, 44(3):355--362,
  1988.

\bibitem{Gor}
Y.~Gordon.
\newblock {\em On Milman's inequality and random subspaces which escape through
  a mesh in $\R^n$}.
\newblock Springer, 1988.

\bibitem{haviv2015restricted}
I.~Haviv and O.~Regev.
\newblock The {R}estricted {I}sometry {P}roperty of subsampled {F}ourier
  matrices.
\newblock {\em arXiv preprint arXiv:1507.01768}, 2015.

\bibitem{johnson1984extensions}
W.~B. Johnson and J.~Lindenstrauss.
\newblock Extensions of {L}ipschitz mappings into a {H}ilbert space.
\newblock {\em Contemporary mathematics}, 26(189-206):1, 1984.

\bibitem{kane2010derandomized}
D.~M. Kane and J.~Nelson.
\newblock A derandomized sparse {J}ohnson-{L}indenstrauss transform.
\newblock {\em arXiv preprint arXiv:1006.3585}, 2010.

\bibitem{klartag2005empirical}
B.~Klartag and S.~Mendelson.
\newblock Empirical processes and random projections.
\newblock {\em Journal of Functional Analysis}, 225(1):229--245, 2005.

\bibitem{Mendel2}
V.~Koltchinskii and S.~Mendelson.
\newblock Bounding the smallest singular value of a random matrix without
  concentration.
\newblock {\em arXiv preprint arXiv:1312.3580}, 2013.

\bibitem{krahmer2011new}
F.~Krahmer and R.~Ward.
\newblock New and improved {J}ohnson-{L}indenstrauss embeddings via the
  {R}estricted {I}sometry {P}roperty.
\newblock {\em SIAM Journal on Mathematical Analysis}, 43(3):1269--1281, 2011.

\bibitem{liberty2011dense}
E.~Liberty, N.~Ailon, and A.~Singer.
\newblock Dense fast random projections and lean {W}alsh {T}ransforms.
\newblock {\em Discrete \& Computational Geometry}, 45(1):34--44, 2011.

\bibitem{Mendel1}
S.~Mendelson.
\newblock Learning without concentration.
\newblock {\em arXiv preprint arXiv:1401.0304}, 2014.

\bibitem{mendelson2007reconstruction}
S.~Mendelson, A.~Pajor, and N.~Tomczak-Jaegermann.
\newblock Reconstruction and subgaussian operators in asymptotic geometric
  analysis.
\newblock {\em Geometric and Functional Analysis}, 17(4):1248--1282, 2007.

\bibitem{nelson2014new}
J.~Nelson, E.~Price, and M.~Wootters.
\newblock New constructions of {RIP} matrices with fast multiplication and
  fewer rows.
\newblock In {\em Proceedings of the Twenty-Fifth Annual ACM-SIAM Symposium on
  Discrete Algorithms}, pages 1515--1528. SIAM, 2014.

\bibitem{companion}
S.~Oymak, B.~Recht, and M.~Soltanolkotabi.
\newblock Sharp time--data tradeoffs for linear inverse problems.
\newblock {\em In preparation.}, 2015.

\bibitem{pilanci2014randomized}
M.~Pilanci and M.~J. Wainwright.
\newblock Randomized sketches of convex programs with sharp guarantees.
\newblock In {\em IEEE International Symposium on Information Theory (ISIT
  2014)}, pages 921--925.

\bibitem{RudelSparse}
M.~Rudelson and R.~Vershynin.
\newblock Sparse reconstruction by convex relaxation: Fourier and gaussian
  measurements.
\newblock In {\em 40th Annual Conference on Information Sciences and Systems},
  pages 207--212, 2006.

\bibitem{talagrand2006generic}
M.~Talagrand.
\newblock {\em The generic chaining: upper and lower bounds of stochastic
  processes}.
\newblock Springer Science \& Business Media, 2006.

\bibitem{TroppConvex}
J.~A. Tropp.
\newblock Convex recovery of a structured signal from independent random linear
  measurements.
\newblock {\em arXiv preprint arXiv:1405.1102}, 2014.

\bibitem{Vers}
R.~Vershynin.
\newblock Introduction to the non-asymptotic analysis of random matrices.
\newblock {\em arXiv preprint arXiv:1011.3027}, 2010.

\bibitem{yap2013stable}
H.~L. Yap, M.~B. Wakin, and C.~J. Rozell.
\newblock Stable manifold embeddings with structured random matrices.
\newblock {\em IEEE Journal on Selected Topics in Signal Processing,},
  7(4):720--730, 2013.

\end{thebibliography}
\bibliographystyle{plain}
\end{document}